\documentclass[a4paper,11pt]{article}
\usepackage{pos}

\usepackage{units}
\usepackage{tikz}
\usepackage{booktabs}
\usepackage{xcolor}
\usepackage{enumitem}
\usepackage{hyperref}
\def\checkmark{\tikz\fill[scale=0.4](0,.35) -- (.25,0) -- (1,.7) -- (.25,.15) -- cycle;}

\newcommand{\quotemark}[1]{``#1''}
\newcommand{\inchsign}{${}^{\prime\prime}$ }

\newenvironment{compressed_itemize}{
\begin{itemize}[topsep=0pt]
  \setlength{\itemsep}{1pt}
  \setlength{\parskip}{0pt}
  \setlength{\parsep}{0pt}
}{\end{itemize}}

\title{The CoMET multiperspective event tracker for wide field-of-view gamma-ray astronomy}
\ShortTitle{The CoMET multiperspective event tracker}

\author*[a]{Ga\v{s}per Kukec Mezek}

\affiliation[a]{Department of Physics and Electrical Engineering, Linnaeus University, 35195 V\"axj\"o, Sweden}

\forColl{CoMET}

\emailAdd{gasper.kukecmezek@lnu.se}

\abstract{The CoMET R\&D project focuses on the development of a new technique for the observation of very high-energy (VHE) $\gamma$-rays from the ground at energies above $\unit[{\sim}200]{GeV}$, thus covering emission from soft-spectrum sources. The CoMET array under study combines 1242 particle detector units, distributed over a circular area of $\unit[{\sim}160]{m}$ in diameter and placed at a very high altitude ($\unit[5.1]{km}$), with atmospheric Cherenkov light detectors.

The atmospheric Cherenkov light detectors, inspired by the \quotemark{HiSCORE} design and improved for the energy range of interest, can be operated together with the particle detectors during clear nights. As such, the instrument becomes a Cosmic Multiperspective Event Tracker (CoMET). CoMET is expected to improve the reconstruction of arrival direction, energy and shower maximum determination for $\gamma$-ray-induced showers during darkness, which is crucial for the reduction of background contamination from cosmic rays. Prototypes of both particle and atmospheric Cherenkov light detectors are already installed at Linnaeus University in Sweden, while in parallel we simulate the full detector response and estimate the reconstruction improvement for $\gamma$-ray events.

In this contribution, we present Monte-Carlo simulations of the detector array, consisting of CORSIKA shower simulations and custom detector response simulations, together with the coupling of particle and atmospheric Cherenkov light information, the reconstruction strategy of the complete array and the detection performance on point-like VHE $\gamma$-ray sources.} 

\FullConference{37$^{\rm{th}}$ International Cosmic Ray Conference (ICRC 2021)\\
		July 12th -- 23rd, 2021\\
		Online -- Berlin, Germany}

\begin{document}
\maketitle

\section{Introduction and aim of the CoMET project}

The CoMET project couples atmospheric shower particle detectors, in the form of water-Cherenkov detectors (WCDs) and scintillation detectors (ALTO project, see \cite{mohan_icrc2021} at this conference) with atmospheric Cherenkov Light Collectors (CLiC).
During darkness, this coupling turns the ALTO array into a \quotemark{Cosmic Multiperspective Event Tracker} (CoMET).
The project, which is currently in its R\&D phase, aims to explore and optimize the technology for a wide-field very-high-energy (VHE) $\gamma$-ray observatory dedicated to extragalactic $\gamma$-ray astronomy. 
CoMET is therefore intended to be installed at high altitudes ($\unit[5.1]{km}$) to optimize the detection of low-energy $\gamma$-rays ($\unit[200]{GeV}$ -- $\unit[100]{TeV}$).
The independent but complementary information gained through the addition of atmospheric Cherenkov light improves the energy resolution and source localization during clear nights.
The latter inherently also reduces background contamination around the direction of $\gamma$-rays from the source and improves the sensitivity for the detection of $\gamma$-ray sources.
%

\section{Coupling particles with atmospheric Cherenkov light during darkness} \label{sec:clic}

\begin{figure}[b] 
    \centering
    \includegraphics[width=0.45\textwidth]{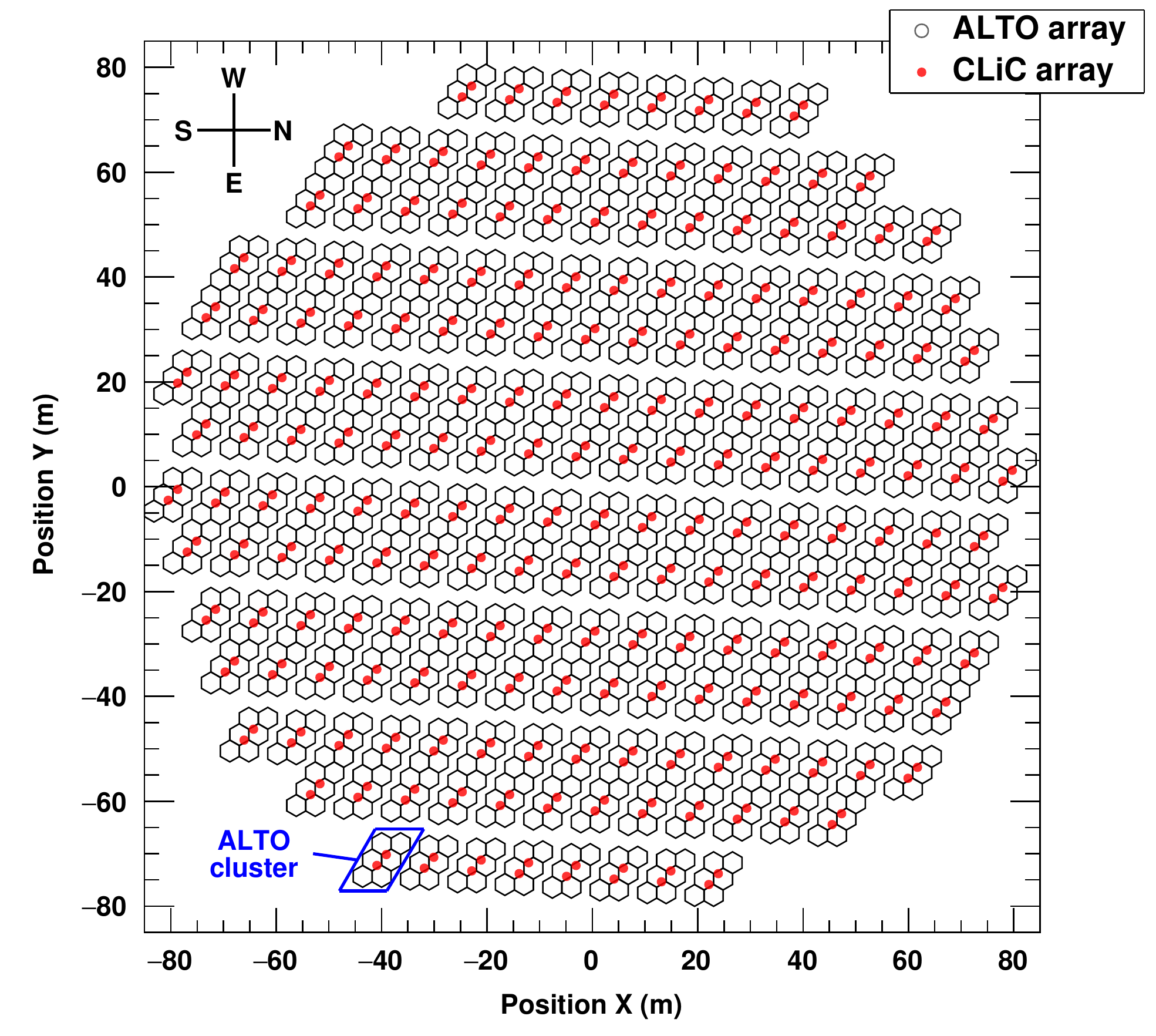}
    \centering
    \includegraphics[width=0.45\textwidth]{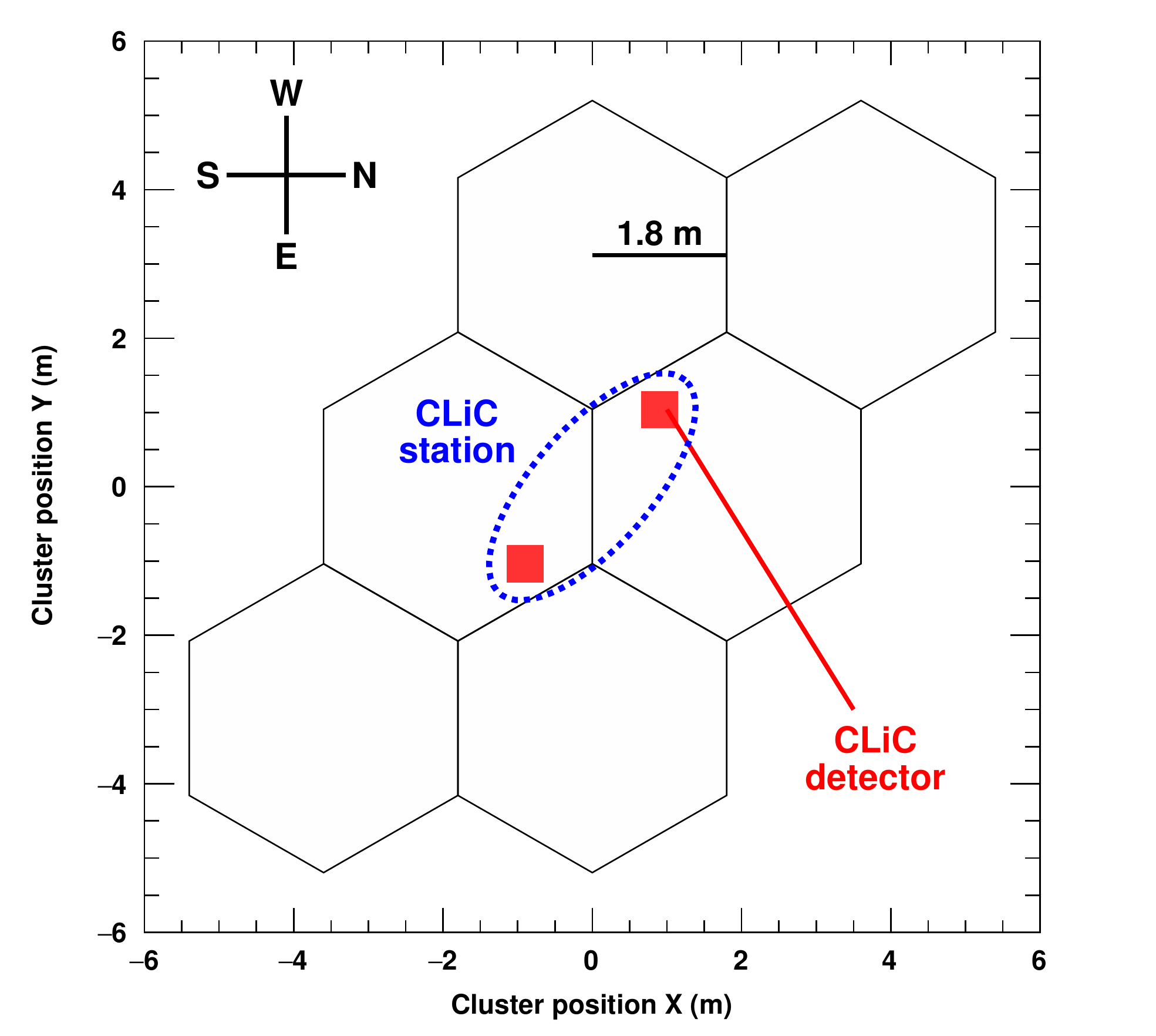}
    \caption{\small{\textbf{Layout of the CoMET array (ALTO and CLiC) and overview of a single array cluster.}
    \emph{Left:} Hexagons mark individual ALTO units consisting of a water-Cherenkov and scintillation detector, while red dots mark individual CLiC detectors.
    \emph{Right:} Magnified view of a single cluster. Each cluster of 6 ALTO units is paired with two CLiC detectors (red squares) that are combined into a CLiC station.
    }}
    \label{fig:altocomet_array}
\end{figure}

%
The proposed design of the ALTO array consists of 1242 particle detector units, distributed over an area of $\unit[\sim 160]{m}$ in diameter, with each cluster combining 6 detector units, see Figure \ref{fig:altocomet_array}.
In the CoMET R\&D project, 414 CLiC detectors are added on top of the ALTO array, with two CLiC detectors per ALTO cluster (called a CLiC station), see Figure \ref{fig:altocomet_array}.

{\bf{CLiC versus HiSCORE.}} The CLiC devices are inspired from the \quotemark{HiSCORE} wide-field-of-view technology adopted in the TAIGA experiment, see \cite{hiscore}, where large Winston cones are coupled with 8\inchsign photomultiplier tubes (PMTs). 
We implement the CLiC technology by instead using smaller 3\inchsign PMTs and combining 8 PMTs per CLiC detector with the aim of improving the signal-to-noise ratio for low-energy atmospheric showers. The signals from two CLiC detectors in a cluster (16 channels in total) are then summed into a single signal.

{\bf{CLiC details.}} The atmospheric Cherenkov light produced by particles in the shower cascade is therefore detected by eight 3\inchsign Hamamatsu R6233 PMTs that constitute one CLiC detector.
To further increase the collection area to $\unit[0.1]{m^2}$ per detector, each PMT is equipped with a 6-sided Winston cone light guide made from aluminized Mylar (85\% reflectivity) and with a $\unit[14]{cm}$ diameter opening, designed to provide an angular cut-off of $30^{\circ}$.
The opening is completely covered by a UV-pass filter with the same transmission characteristics as Schott UG11 glass (ZWB1 from \cite{bandpass_filter}) to filter out the night sky background (NSB) and ambient light (especially at the prototype site). All PMTs are considered to have an operational gain of $2\times 10^6$.

{\bf{Trigger strategy.}} The addition of CLiC detectors to the existing ALTO array does not change our triggering strategy, which is still based on the coincidence of detectors in an ALTO cluster and then reading out CLiC stations connected to a triggered cluster with signals above the $\unit[4]{mV}$ threshold. To increase the number of triggered CLiC stations for each event, our current strategy for simulations also includes reading of CLiC stations, not connected to a triggered cluster, but with signals above the $\unit[8]{mV}$ threshold.

\section{Atmospheric Cherenkov light simulations} \label{sec:simulation}

{\bf{Atmospheric shower simulations}.} We performed air shower simulations using the CORSIKA package (version 7.4387) \cite{corsika}, as described in \cite{semla_paper}, but with added atmospheric Cherenkov light emission according to the Bernlöhr model \cite{bernlohr}.

With already existing simulations for the ALTO array, we had to ensure that it would be possible to include the new CLiC related information without repeating the time-consuming GEANT4 \cite{geant4} simulations.
Since CLiC stations are triggered by their corresponding ALTO clusters, we made sure that each simulated shower was exactly the same on the particle side, but with added atmospheric Cherenkov light information. This led to a number of CORSIKA modifications, including larger step sizes during Cherenkov light emission, treatment of particles below their energy cut-off and implementation of a separate random number generator to keep all original random numbers unchanged.
These modifications affected the number of photo-electrons seen by the CLiCs, with the largest differences observed close to the shower core. On average, we tend to get additional photons at low primary energies and a photon deficiency at high energies.
For $\gamma$-rays the maximum difference is $< 6\%$, while for protons it is $< 12\%$ over the energy range of CoMET. This mismatch caused by CORSIKA modifications will be studied in greater detail in a future study, and subsequently corrected during the CLiC detector response step.

\begin{figure}[t] 
    \centering
    \includegraphics[width=0.584\textwidth]{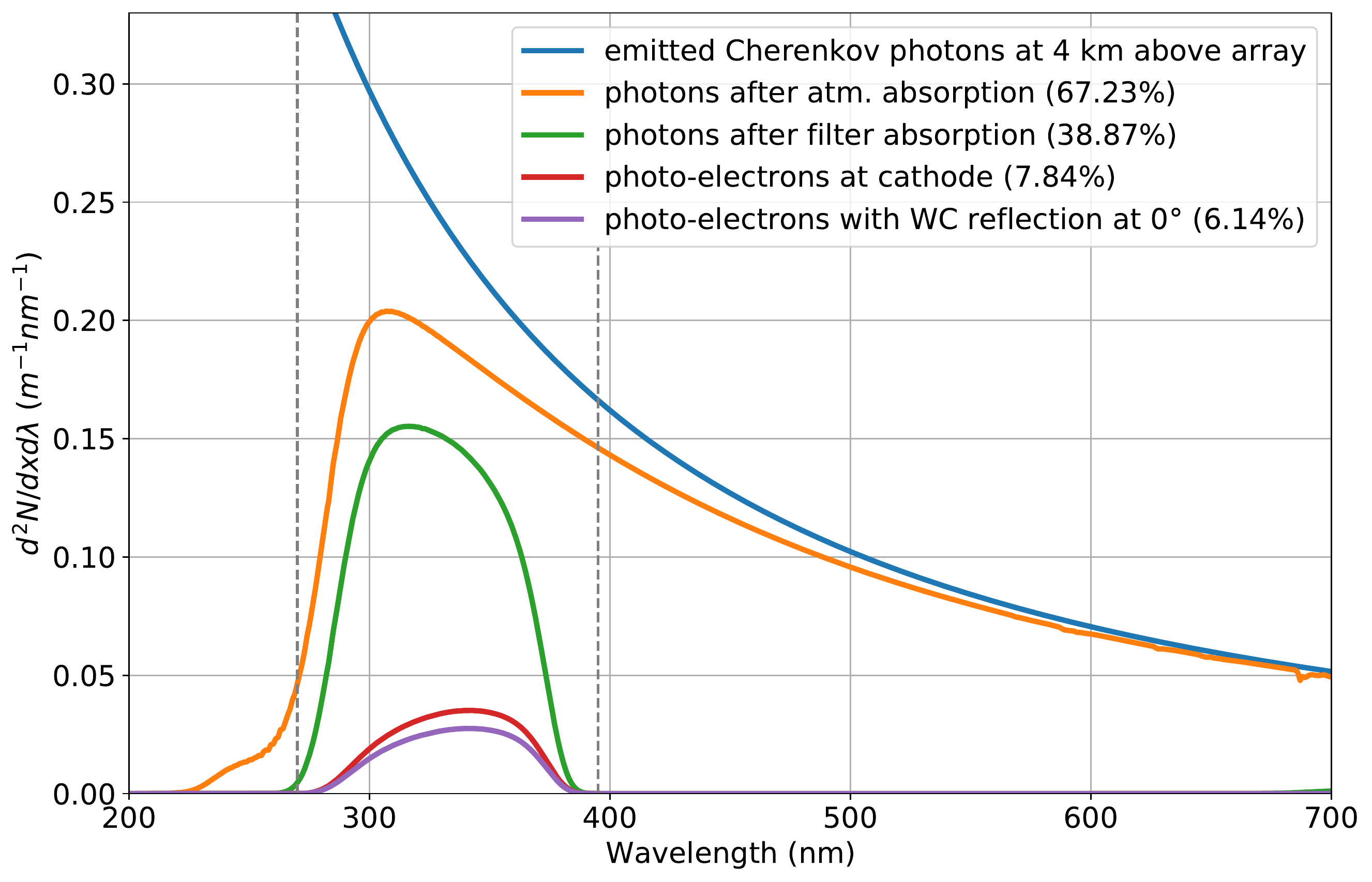}
    \centering
    \includegraphics[width=0.408\textwidth]{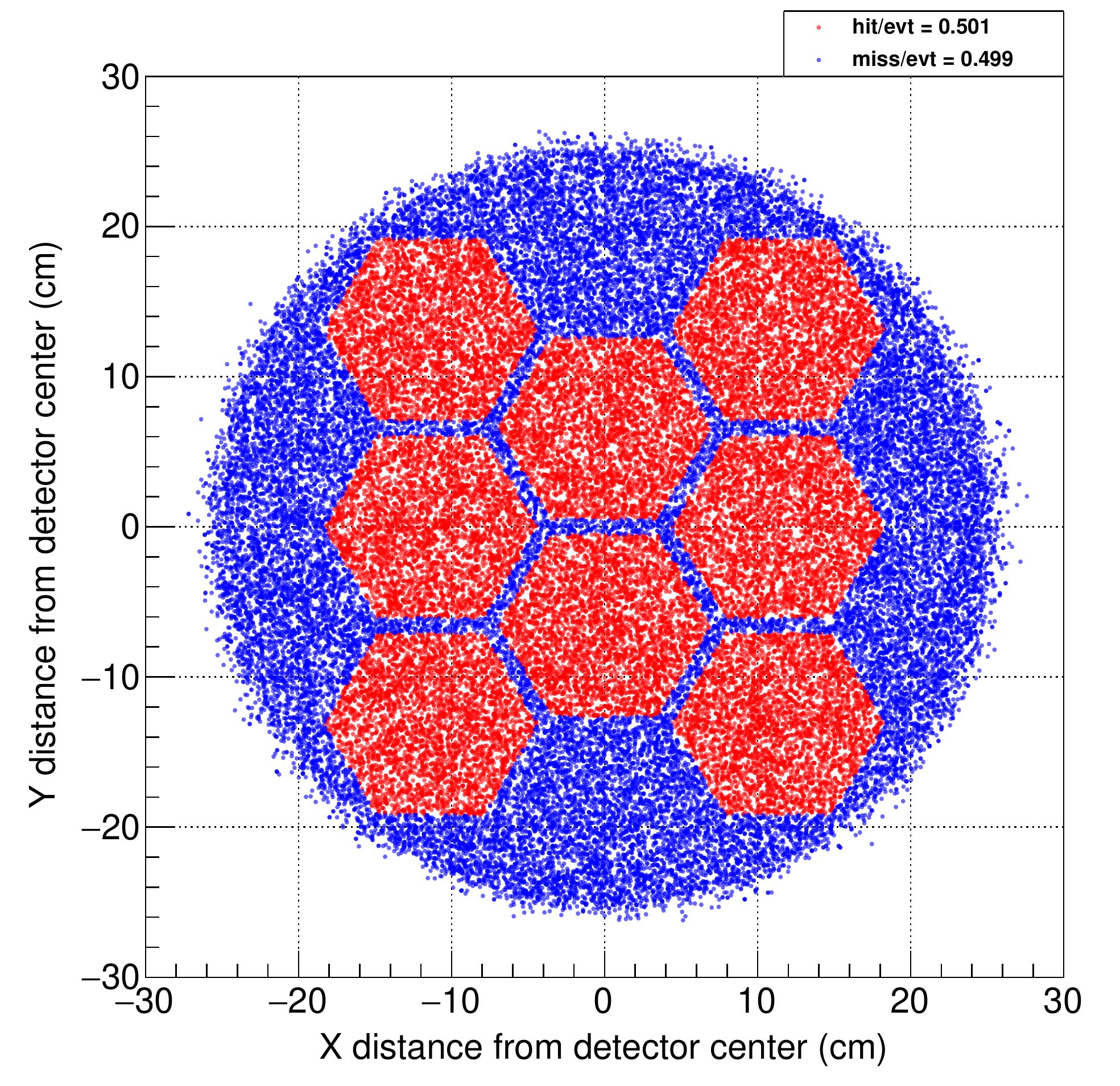}
    \caption{\small{\textbf{Transmission of atmospheric Cherenkov light and layout of one CLiC detector.}
    \emph{Left:} Spectrum of atmospheric Cherenkov light before and after transmission through the complete detector simulation. The gray dashed vertical lines denote limits for random photon wavelength assignment after the CORSIKA simulation. The emission height is selected to be at $\unit[4]{km}$ above the array, which coincides with the mean shower maximum for $\unit[1]{TeV}$ $\gamma$-rays.
    \emph{Right:} CLiC detector layout shown with hit positions from a simplified ray-tracing simulation.
    }}
    \label{fig:clic_detector}
\end{figure}

{\bf{CLiC detectors response}.} After the CORSIKA shower simulation, the emitted Cherenkov photons have to be followed from their emission point until they produce a signal in one of the CLiC PMTs. For this treatment, we produced a detector simulation code, which is based on the \texttt{sim\_telarray} simulation package \cite{bernlohr}. However, since we do not require exact positional precision, as needed by imaging atmospheric Cherenkov telescopes, we could greatly simplify the ray-tracing component. The CLiC simulation code handles the following aspects of atmospheric Cherenkov light transmission and detector response, see Figure \ref{fig:clic_detector}:
\begin{compressed_itemize}
    \item Random assignment of wavelength according to the Cherenkov light spectrum.
    \item Transmission through the atmosphere, using a MODTRAN \cite{modtran} atmospheric transmittance model for a desert tropical atmosphere.
    \item Transmission through the UV-pass filter.
    \item Winston cone simplified ray-tracing, by selecting only photons falling onto the top opening, see Figure \ref{fig:clic_detector} (right panel).
    \item Reflection and angular efficiency on the walls of the Winston Cone. 
    \item PMT quantum efficiency for converting photons into photo-electrons.
    \item Conversion of the photo-electron timing waveform for each CLiC detector into a signal, and summation of signals to produce the response of a CLiC station.
\end{compressed_itemize}

\section{CoMET air-shower reconstruction using ALTO and CLiC stations} \label{sec:reconstruction}

During the event reconstruction stage, we took the existing shower core estimation from the ALTO array and kept it fixed, since the particle detectors have finer granularity and so give a better estimate. The main aim for including CLiC stations is to extract independent information related to the shower development and, in combination with ALTO, improve the $\gamma$-ray source localisation.

\begin{figure}[t]
    \centering
    \includegraphics[width=0.46\textwidth]{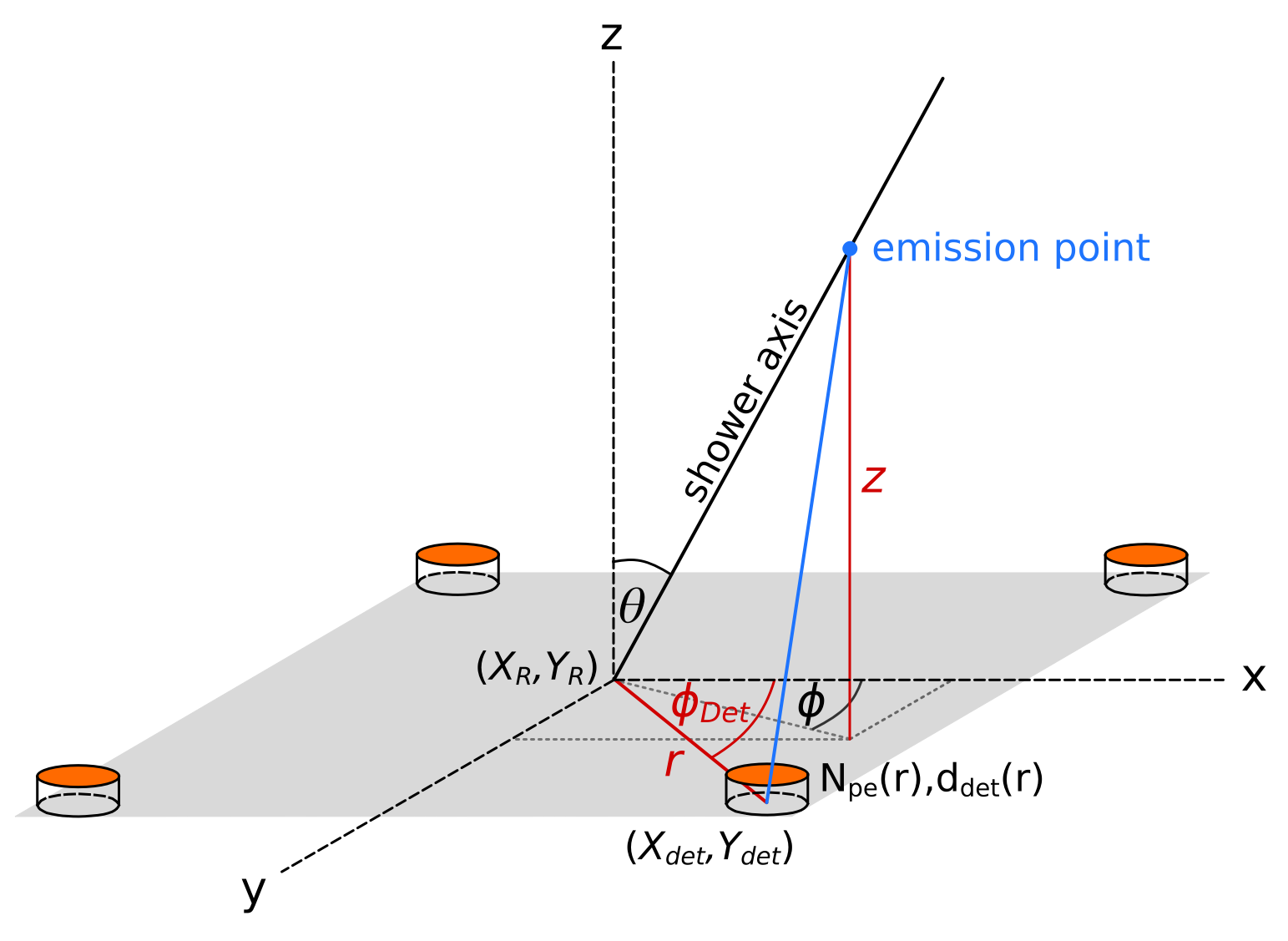}
    \centering
    \includegraphics[width=0.46\textwidth]{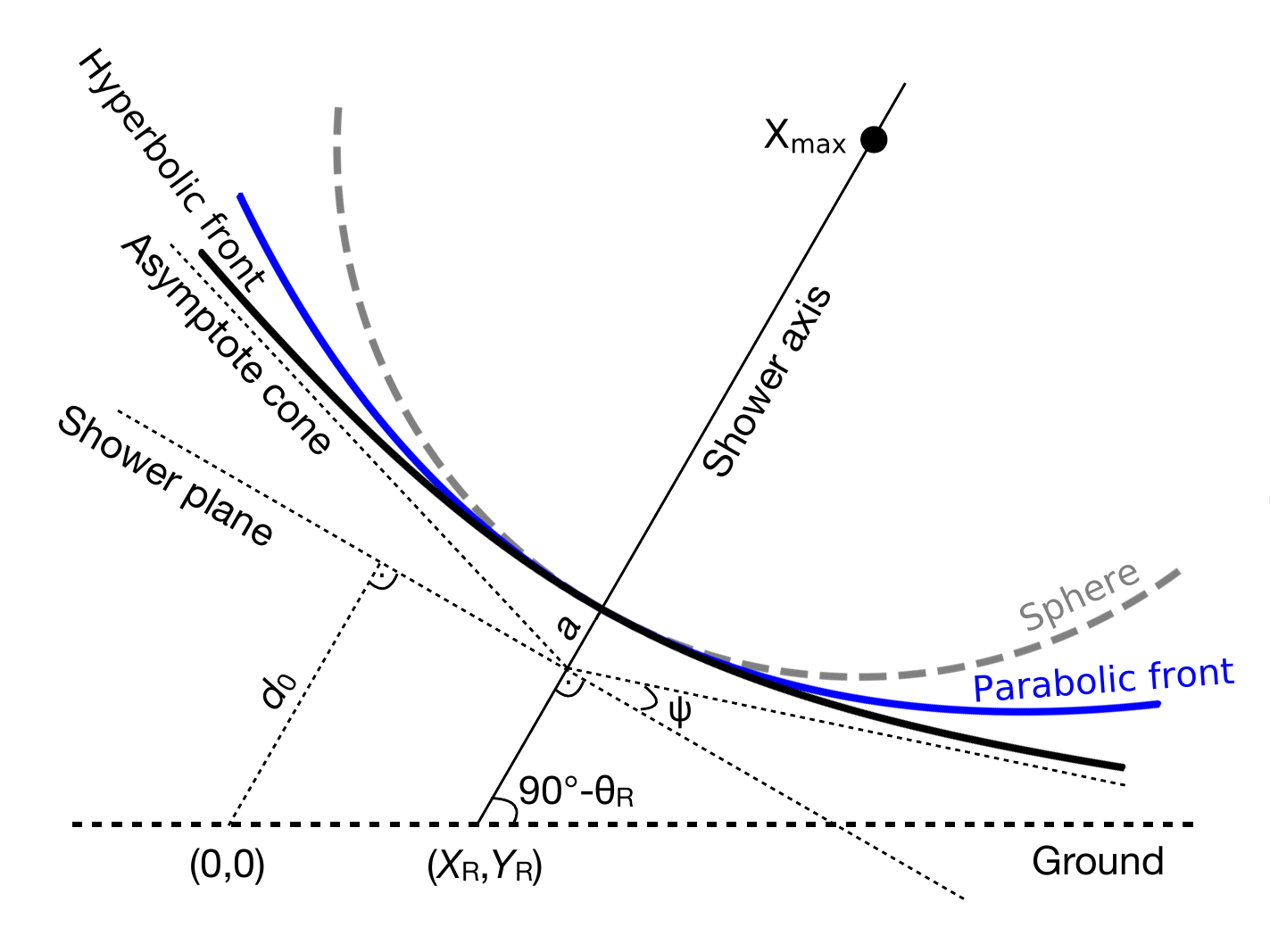}
    \caption{\small{\textbf{Representation of variables and shower fronts used during the reconstruction.}
    \emph{Left:} Graphical representation for emission and transmission of atmospheric Cherenkov light (blue line) and variables used during the reconstruction.
    \emph{Right:} Graphical representation of the different shower front profiles. Hyperbolic and parabolic fronts describing particles and atmospheric Cherenkov light, respectively, are used in the timing and combined fits (adapted from \cite{semla_paper}).
    }}
    \label{fig:fitting_explain}
\end{figure}

The complete procedure for CoMET includes a number of fits over the integrated charge and timing information gathered by each CLiC station. A detailed explanation of each of these fits will be outlined in a forthcoming paper, but the relevant fitting variables are listed in Table \ref{table:reco_details}, and a representation of variables and shower fronts used during the reconstruction are shown in Figure \ref{fig:fitting_explain}. For all fits described in this reconstruction loop, we require at least 3 triggered ALTO array detectors and 3 CLiC stations above their signal threshold.
\begin{table}[t]
    \centering
    {\footnotesize
    \begin{tabular}{lp{0.09\textwidth}p{0.40\textwidth}ccc}
        \toprule
        \textbf{Variable} & \textbf{Used in} & \textbf{Definition} &\textbf{LDF} & \textbf{Parabolic} & \textbf{Combined}\\ 
        & \textbf{SEMLA} &  &\textbf{fit} & \textbf{timing fit} & \textbf{timing fit}\\ 
        \midrule
        $X_{\rm R} \pm \Delta X_{\rm R}$ & \checkmark & Core position (X-coordinate) & fixed & fixed & fixed \\ 
        $Y_{\rm R} \pm \Delta Y_{\rm R}$ & \checkmark & Core position (Y-coordinate) & fixed & fixed & fixed \\ 
        \midrule
        $N_{\rm 0} \pm \Delta N_{\rm 0} $ &  & Normalization factor of the LDF fit & free &  &  \\ 
        $p_{\rm R} \pm \Delta p_{\rm R}$ & \checkmark & LDF slope parameter & free &  &  \\ 
        \midrule
        $d_{\rm 0H} \pm \Delta d_{\rm oH} $ &  & Perpendicular distance between the shower plane and the origin of detector coordinates (CLiC) &  & free & free \\ 
        $z_{\rm R} \pm \Delta z_{\rm R}$ &  & Vertical height of Cherenkov light emission &  & free & free \\ 
        \midrule
        $\phi_{\rm R} \pm \Delta \phi_{\rm R}$ & \checkmark & Azimuth angle & fixed & fixed & free  \\ 
        $\theta_{\rm R} \pm \Delta \theta_{\rm R}$ & \checkmark & Zenith angle & fixed & fixed & free \\ 
        $d_{\rm 0A} \pm \Delta d_{\rm 0A} $ &  & Perpendicular distance between the shower plane and the origin of detector coordinates (ALTO) & &  &  free \\ 
        $\psi \pm \Delta \psi$ &  & Cone angle of the asymptote with respect to the shower plane & & & free \\ 
        $a \pm \Delta a$ &  & Perpendicular distance of the hyperbola vertex from the shower plane & &  & free \\ 
        \bottomrule
    \end{tabular}
    }
    \caption{\textbf{Variables from the shower reconstruction loop.}
        Variables with the check marks in the second column are used in the SEMLA analysis \cite{semla_paper} for CoMET. The prefix $\Delta$ represents the error associated with the fitted variable. Core position is given a fixed value from the ALTO reconstruction for all fits.
    }
    \label{table:reco_details}
\end{table}

{\bf{CLiC array reconstruction.}} The density of atmospheric Cherenkov photons with respect to the distance from the shower core in the shower plane is defined by the lateral distribution function (LDF). While the LDF generally has an exponential behaviour up to a distance of $\sim \unit[120]{m}$ and a power-law behaviour beyond that \cite{ldf_blanca,ldf_fit_tunka}, the CLiCs only observe light produced in the exponential part due to the size of the CoMET array. The LDF fit over the integrated charges of CLiC stations is not used further in the reconstruction chain, but it does include shower-development-sensitive observables which can be used in the subsequent analysis, such as the exponential LDF slope $p_{\rm R}$.\\
If we consider the distribution of signal peak times of atmospheric Cherenkov light with respect to the position of each CLiC station in the array, then we are able to use a parabolic fitting function to reconstruct the arrival direction of the primary. This parabolic timing model has already been used by the HiSCORE detector, as described in \cite{hiscore_reco}, and it works equally well for our purpose. However, at this stage, the function is only used to perform initial estimations of its two free parameters, as seen in Table \ref{table:reco_details}, which are then later used in the combined timing fit.

{\bf{Combined CoMET reconstruction.}} The final fitting step combines the information from the detection of particles and atmospheric Cherenkov light. Based on the separate timing fits of ALTO (hyperbolic fit \cite{hyperbolic1}) and CLiC stations (parabolic fit), we aim to improve the final arrival direction reconstruction. This is done through a final $\chi^2$-minimization by combining the respective performance of each fit as $\chi^2 = \chi_{\rm hyper}^2 + \chi_{\rm parab}^2$, and keeping all but the core position as a bound free parameter.

\section{Preliminary sensitivity of CoMET during darkness} \label{sec:prelimResults}

A preliminary analysis, in order to evaluate the importance of adding atmospheric Cherenkov light information, has been performed following the SEMLA analysis structure, consisting of four subsequent stages, see \cite{semla_paper}.
Using this analysis procedure, we additionally extracted four new observables to be used in the machine learning processes. 
For cleaning badly-reconstructed events in Stage B, we used the $\chi^2_{\rm parab}$ value of the parabolic fit and the LDF slope parameter $p_{\rm R}$. 
For $\gamma$/hadron separation in Stage C, we added $p_{\rm R}$ and the LDF function value at $\unit[60]{m}$ from the shower core $p_{60}$.
In Stage D, aimed at energy reconstruction, we used $p_{60}$ and the total number of photo-electrons detected in the CLiC array $N_{\rm pe}$, both showing good correlation with primary energy.

\begin{figure}[b] 
    \centering
    \includegraphics[width=0.95\textwidth]{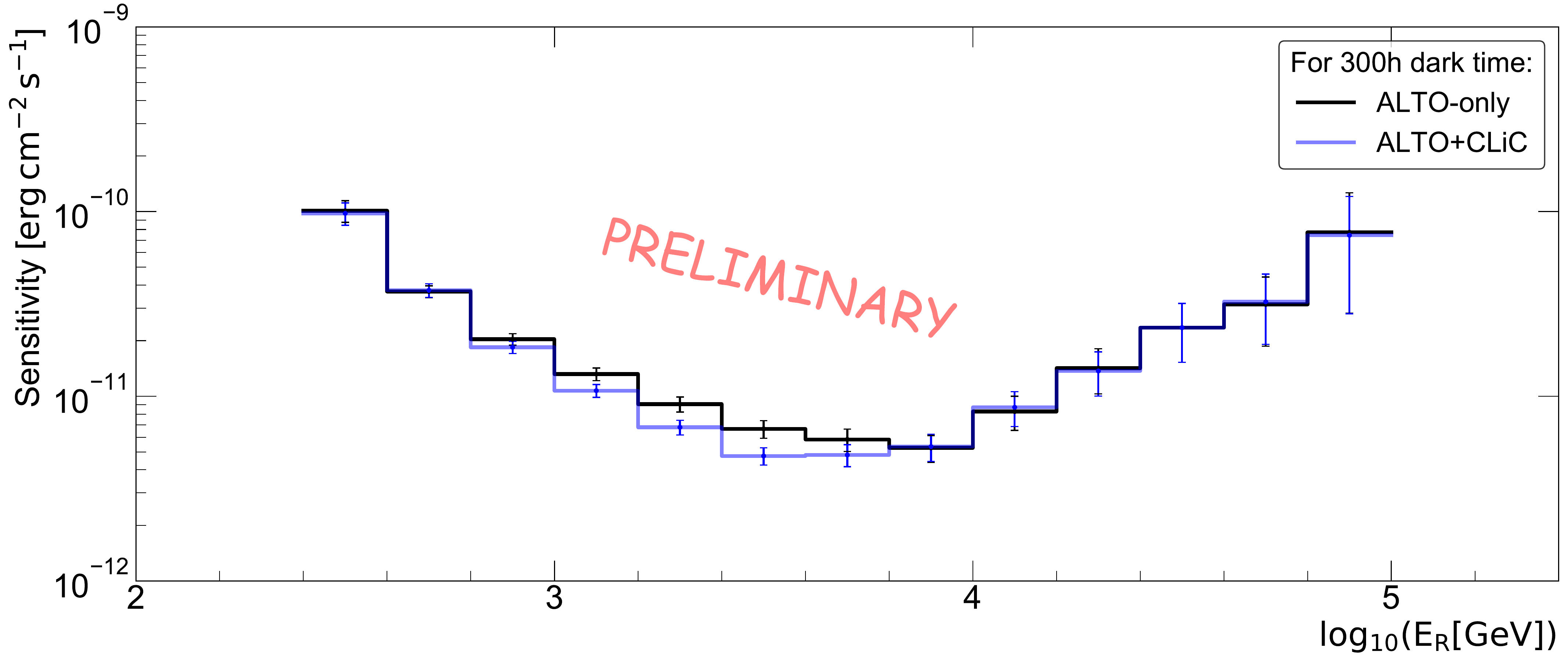}
    \caption{\small{\textbf{Improvement in sensitivity by adding CLiC detectors to the ALTO array.}
    Expected CoMET sensitivity for a point-like $\gamma$-ray source described by a power-law spectral index of 2, at a zenith of $18^{\circ}$ for 300 hours (during dark-time), compared to the ALTO-only performance (\emph{config-Q2} from \cite{mohan_icrc2021}).
    }}
    \label{fig:semla_performance}
\end{figure}

The preliminary sensitivity of CoMET during darkness, see Figure \ref{fig:semla_performance}, shows that we can achieve a clear improvement in the energy range between $\unit[600]{GeV}$ and $\unit[6]{TeV}$ when compared to the ALTO-only analysis (\emph{config-Q2} from \cite{mohan_icrc2021}). 
The improvement during darkness by the addition of the CLiC detectors results in a 10\% better angular resolution in the full energy range, a 30\% better energy resolution at $\unit[1]{TeV}$ and a 12\% better background suppression with respect to the performance of the ALTO array, at the expense of only a 1\% loss of $\gamma$-ray events.    

\section{CoMET prototype activities} \label{sec:prototype}

After almost two years of gathering data with two ALTO unit prototypes installed at the Linnaeus University campus in Växjö, Sweden ($\unit[160]{m}$ a.s.l.), we installed the first CLiC pre-prototype in late September 2020, featuring four Hamamatsu R6233 PMTs with remotely controlled HV-generating sockets. It is located on top of one of the WCDs, in a box with a remotely-controlled shutter and a heating wire system that prevents freezing or fogging during cold nights. A continuous monitoring of cloud coverage and sky brightness allows to automatically handle the shutter as well as the PMT high voltage according to the sky conditions. The signals from the four PMTs are summed using an analog summation card, and the resulting signal is sent to one channel of the ALTO prototype readout electronics: a 16-channel WaveCatcher \cite{wavecatcher}. 
%

In addition to the CLiC pre-prototype, we installed four stations each featuring an 8\inchsign Electron Tube KB9352 PMT, with a $\times 10$ amplifier and one Winston cone, derived from HiSCORE \cite{hiscore} (henceforth denoted as mini-HiSCORE).


{\bf{First prototype observation results.}} Earlier measurements in Autumn--Winter 2020--2021, with the pre-prototype and mini-HiSCOREs, have demonstrated the feasibility of the concept, since a Cherenkov signal from atmospheric showers has been detected in correlation with particle detector signals. Although Cherenkov light is induced by particles passing through the filter and PMT glass, the measurements with an open shutter have shown a detection enhancement rate of a factor of $\approx 30$, compared to observations with a closed shutter. The effectiveness of the UV filters allowed to keep a gain of $2\times 10^6$ for the pre-prototype during moonless nights, even in the vicinity of a city, provided that the sky brightness is darker than 19 mag/arcsec$^2$. These filters may not be necessary for a dark site ($\approx 22$ mag/arcsec$^2$).

{\bf{Future prototype activities.}} During Summer 2021, we plan to complete the installation of a full prototype of one CLiC detector, with eight R6233 PMTs. Observations will then be conducted in Autumn--Winter 2021--2022, together with the pre-prototype and the four mini-HiSCORE detectors. Using all atmospheric Cherenkov light detectors, we aim to measure the Cherenkov light-front direction, which is analogous to the arrival direction measurement from ALTO particle detectors. The performance of both will be estimated and compared in order to study the arrival direction and timing accuracy of the CLiC prototype. The event rate of this prototype, which relates to its effective collection area, will also be studied, in addition to its behaviour in various sky brightness conditions.

\section{Conclusion and future activities} \label{sec:results}

In this contribution, we present a preliminary study on the possibility to add atmospheric Cherenkov light detectors to a particle detector array during darkness for the purpose of ($\unit[200]{GeV}$ -- $\unit[100]{TeV}$) ground-based and wide field-of-view $\gamma$-ray astronomy. 
The added information results in a better angular and energy resolution, and an improved background suppression, at the expense of just a 1\% loss of $\gamma$-rays. 
The performance during darkness is improved in the range [$\unit[600]{GeV}$, $\unit[6]{TeV}$] with our relatively simple CLiC technology. 
The results shown here are for a minimum of 8 WCDs and 3 CLiC stations.  


The future activities concerning the analysis will be focused on the possibility of improving the sensitivity for $E<\unit[600]{GeV}$, by studying the effect of lowering the minimum number of WCDs considered in the analysis. 
On the technical side, future improvements concerning the analysis include the correction of the Cherenkov photon response on the ground in simulations, as mentioned in Section \ref{sec:simulation}, the study of the CLiC array response at different altitudes and a maximum-likelihood reconstruction procedure. 
The prototyping effort will concentrate on the possibility of achieving a better timing accuracy and summation through the operation of two CLiC detectors. 

\acknowledgments
\href{https://alto-gamma-ray-observatory.org/acknowledgements/}{https://alto-gamma-ray-observatory.org/acknowledgements/}


\clearpage
\section*{Full Authors List: \Coll\ Collaboration}
%
%
\scriptsize
\noindent
Yvonne Becherini$^1$,
Tomas Bylund$^1$,
Jean-Pierre Ernenwein$^2$,
Gašper Kukec Mezek$^1$,
Michael Punch$^3$,
Patrizia Romano$^4$,
Ahmed Saleh$^5$,
Mohanraj Senniappan$^1$,
Satyendra Thoudam$^5$,
Martin Tluczykont$^6$,
Stefano Vercellone$^4$ \\

\noindent
$^1$Department of Physics and Electrical Engineering, Linnaeus University, 35195 Växjö, Sweden.
$^2$Aix Marseille Univ, CNRS/IN2P3, CPPM, Marseille, France.
$^3$Universit\'e de Paris, CNRS, Astroparticule et Cosmologie, F-75013 Paris, France
$^4$INAF-Osservatorio Astronomico di Brera, Brera, Italy.
$^5$Department of Physics, Khalifa University, PO Box 127788, Abu Dhabi, United Arab Emirates.
$^6$Institut für Experimentalphysik, Universität Hamburg, Hamburg, Germany.


\begin{thebibliography}{99}
\bibitem{mohan_icrc2021}
M. Senniappan et al., \emph{Expected performance of the ALTO particle detector array designed for 200 GeV -- 50 TeV gamma-ray astronomy}, at this conference.
\vspace{-2.5mm}

\bibitem{hiscore}
M. Tluczykont et al., \emph{The HiSCORE concept for gamma-ray and cosmic-ray astrophysics beyon 10 TeV}, \href{https://doi.org/10.1016/j.astropartphys.2014.03.004}{Astropart. Phys. \textbf{56} (2014) 42 -- 53}.
\vspace{-2.5mm}

\bibitem{bandpass_filter}
Shijiazhuang Tangsinuo Optoelectronic Technology,  
\href{http://www.tangsinuo.com}{http://www.tangsinuo.com} (accessed in June 2021).
\vspace{-2.5mm}

\bibitem{corsika}
D. Heck et al., \emph{CORSIKA: A Monte Carlo code to simulate extensive air showers}, FZKA 6019 (1998).
\vspace{-2.5mm}

\bibitem{semla_paper}
M. Senniappan et al., \emph{Signal extraction in atmospheric shower arrays designed for $\unit[200]{GeV} - \unit[50]{TeV}$ $\gamma$-ray astronomy}, \href{https://arxiv.org/abs/2105.06728}{\texttt{[astro-ph.IM/2105.06728]}}, accepted by JINST.
\vspace{-2.5mm}

\bibitem{bernlohr}
K. Bernlöhr, \emph{Simulation of Imaging Atmospheric Cherenkov Telescopes with CORSIKA and sim\_telarray}, \href{https://doi.org/10.1016/j.astropartphys.2008.07.009}{Astropart. Phys. \textbf{30} (2008) 149}.
\vspace{-2.5mm}

\bibitem{geant4}
S. Agostinelli et al., \emph{Geant4 — a simulation toolkit}, CERN-IT-2002-003.
\vspace{-2.5mm}

\bibitem{modtran}
F. X. Kneizys et al., \emph{The MODTRAN 2/3 Report and LOWTRAN 7 model}, Phillips Laboratory, Hanscom AFB, MA 01731, USA (1996). 
\vspace{-2.5mm}

\bibitem{ldf_blanca}
J.W. Fowler et al., \emph{A Measurement of the Cosmic Ray Spectrum and Composition at the Knee}, \href{https://doi.org/10.1016/S0927-6505(00)00139-0}{Astropart. Phys. \textbf{15} (2001) 49 -- 64}.
\vspace{-2.5mm}

\bibitem{ldf_fit_tunka}
N. M. Budnev et al., \emph{Cosmic Ray Energy Spectrum and Mass Composition from $10^{15}$ to $\unit[10^{17}]{eV}$ by Data of the Tunka EAS Cherenkov Array}, 29th ICRC, Pune \textbf{00} (2005) 101 -- 106 \href{https://arxiv.org/abs/astro-ph/0511215}{\texttt{[astro-ph/0511215]}}.
\vspace{-2.5mm}

\bibitem{hiscore_reco}
D. Hampf, M. Tluczykont, Dieter Horns \emph{Event reconstruction techniques for the wide-angle air Cherenkov detector HiSCORE}, \href{https://doi.org/10.1016/j.nima.2013.02.016}{Nucl. Instrum. Meth. A \textbf{712} (2013) 137 -- 146}.
\vspace{-2.5mm}
%

\bibitem{hyperbolic1}
A. Corstanje et al., \emph{The shape of the radio wavefront of extensive air showers as measured with LOFAR}, \href{https://doi.org/10.1016/j.astropartphys.2014.06.001}{Astropart. Phys. \textbf{61} (2015) 22}.
\vspace{-2.5mm}

\bibitem{wavecatcher}
D. Breton et al.,  
\emph{The WaveCatcher family of SCA-based 12-bit 3.2-GS/s fast digitizers}, \href{https://doi.org/10.1109/RTC.2014.7097545}{19th IEEE-NPSS Real Time Conference (2014) 1 -- 8}.
\vspace{-2.5mm}

\end{thebibliography}
\end{document}